\documentclass[conference, a4paper]{IEEEtran}
\IEEEoverridecommandlockouts
\usepackage{cite}
\usepackage{amsmath,amssymb,amsfonts}
\usepackage{algorithmic}
\usepackage{graphicx}
\usepackage{textcomp}
\usepackage{xcolor}
\usepackage{paralist}
\usepackage{subcaption}
\usepackage{xspace}
\usepackage{hyperref}

\usepackage[export]{adjustbox}
\usepackage{eso-pic}
\usepackage[most]{tcolorbox}

\usepackage{hyperxmp}
\hypersetup{
    pdfauthor={Martin Kaiser, Rene Griessl, Nils Kucza, Carola Haumann, Lennart Tigges, Kevin Mika, Jens Hagemeyer, Florian Porrmann, Ulrich Rückert, Micha vor dem Berge, Stefan. Krupop, Mario Porrmann, Marco Tassemeier, Pedro Trancoso, Fareed Quararyah, Stavroula Zouzoula, Antonio Casimiro, Alysson Bessani, Jose Cecilio, Stefan Andersson, Oliver Brunnegard, Olof Eriksson, Roland Weiss, Franz Meierhöfer, Hans Salomonsson, Elaheh Malekzadeh, Daniel Ödman, Anum Khurshid, Pascal Felber, Marcelo Pasin, Valerio Schiavoni, Jämes Ménétrey, Karol Gugula, Piotr Zierhoffer, Eric Knauss, Hans-Martin Heyn},
    pdftitle={VEDLIoT: Very Efficient Deep Learning in IoT},
    pdflicenseurl={https://doi.org/10.23919/DATE54114.2022.9774653}
}

\hypersetup{
plainpages=false,
hypertexnames=false,
colorlinks=true,
colorlinks=true,
anchorcolor=black,
citecolor=black,
filecolor=black,
linkcolor=black,
menucolor=black,
urlcolor=black,
pdfpagemode=UseOutlines,
bookmarksopen=true,
bookmarksnumbered=true,
pdfstartpage={1},
pdftitle={VEDLIoT: Very Efficient Deep Learning in IoT},
pdfauthor={Martin Kaiser, Rene Griessl, Nils Kucza, Carola Haumann, Lennart Tigges, Kevin Mika, Jens Hagemeyer, Florian Porrmann, Ulrich Rückert, Micha vor dem Berge, Stefan. Krupop, Mario Porrmann, Marco Tassemeier, Pedro Trancoso, Fareed Quararyah, Stavroula Zouzoula, Antonio Casimiro, Alysson Bessani, Jose Cecilio, Stefan Andersson, Oliver Brunnegard, Olof Eriksson, Roland Weiss, Franz Meierhöfer, Hans Salomonsson, Elaheh Malekzadeh, Daniel Ödman, Anum Khurshid, Pascal Felber, Marcelo Pasin, Valerio Schiavoni, Jämes Ménétrey, Karol Gugula, Piotr Zierhoffer, Eric Knauss, Hans-Martin Heyn},
pdfsubject={The VEDLIoT project targets the development of energy-efficient Deep Learning for distributed AIoT applications. A holistic approach is used to optimize algorithms while also dealing with safety and security challenges. The approach is based on a modular and scalable Cognitive IoT hardware platform. Using modular microserver technology enables the user to configure the hardware to satisfy a wide range of applications. VEDLIoT offers a complete design flow for Next-Generation IoT devices required for collaboratively solving complex Deep Learning applications across distributed systems. The methods are tested on various use-cases ranging from Smart Home to Automotive and Industrial IoT appliances. VEDLIoT is an H2020 EU project which started in November 2020. It is currently in an intermediate stage with the first results available.},
pdfkeywords={IoT, AIoT, Microserver, Deep-Learning, Edge-Computing, FPGA, Accelerator, Energy-Efficiency}
} 
\usepackage{siunitx}
\usepackage{placeins}
\usepackage{csquotes}
\usepackage{listings}

\definecolor{mGreen}{rgb}{0,0.6,0}
\definecolor{mGray}{rgb}{0.5,0.5,0.5}
\definecolor{mPurple}{rgb}{0.58,0,0.82}
\definecolor{backgroundColour}{rgb}{0.95,0.95,0.92}

\lstdefinestyle{CStyle}{
	basicstyle=\tiny,
	escapeinside={<@}{@>},
    backgroundcolor=\color{backgroundColour},   
    commentstyle=\color{mGreen},
    keywordstyle=\color{magenta},
    numberstyle=\tiny\color{mGray},
    stringstyle=\color{mPurple},
    breakatwhitespace=false,         
    breaklines=true,                 
    captionpos=b,                    
    keepspaces=true,                 
    numbers=left,                    
    numbersep=5pt,                  
    showspaces=false,                
    showstringspaces=false,
    showtabs=false,                  
    tabsize=2,
    language=C
}

\def\BibTeX{{\rm B\kern-.05em{\sc i\kern-.025em b}\kern-.08em
    T\kern-.1667em\lower.7ex\hbox{E}\kern-.125emX}}
\makeatletter\begin{document}

\title{VEDLIoT: Very Efficient Deep Learning in IoT
\thanks{This publication incorporates results from the VEDLIoT project, which received funding from the European Union's Horizon 2020 research and innovation programme under grant agreement No 957197.}
}

\author{
  \IEEEauthorblockN{\href{mailto:mkaiser@techfak.uni-bielefeld.de}{M.~Kaiser}\IEEEauthorrefmark{1}, \href{mailto:rgriessl@techfak.uni-bielefeld.de}{R.~Griessl}\IEEEauthorrefmark{1}, \href{mailto:nkucza@techfak.uni-bielefeld.de}{N.~Kucza}\IEEEauthorrefmark{1},  \href{mailto:chaumann@cor-lab.uni-bielefeld.de}{C.~Haumann}\IEEEauthorrefmark{1}, \href{mailto:ltigges@techfak.uni-bielefeld.de}{L.~Tigges}\IEEEauthorrefmark{1}, \href{mailto:kmika@techfak.uni-bielefeld.de}{K.~Mika}\IEEEauthorrefmark{1}, \href{mailto:jhagemey@techfak.uni-bielefeld.de}{J.~Hagemeyer}\IEEEauthorrefmark{1},  \href{mailto:fporrmann@techfak.uni-bielefeld.de}{F.~Porrmann}\IEEEauthorrefmark{1},\\ \href{mailto:rueckert@techfak.uni-bielefeld.de}{U.~R\"uckert}\IEEEauthorrefmark{1},
  \href{mailto:micha.vordemberge@christmann.info}{M.~vor~dem~Berge}\IEEEauthorrefmark{2}, \href{mailto:Stefan.Krupop@christmann.info}{S.~Krupop}\IEEEauthorrefmark{2},
  \href{mailto:mporrmann@uni-osnabrueck.de}{M.~Porrmann}\IEEEauthorrefmark{3},
  \href{mailto:marco.tassemeier@uni-osnabrueck.de}{M.~Tassemeier}\IEEEauthorrefmark{3}, 
  \href{mailto:ppedro@chalmers.se}{P.~Trancoso}\IEEEauthorrefmark{4},
  \href{mailto:qarayah@chalmers.se}{F.~Qararyah}\IEEEauthorrefmark{4}, \\
  \href{mailto:zouzoula@chalmers.se}{S.~Zouzoula}\IEEEauthorrefmark{4}
  \href{mailto:casim@ciencias.ulisboa.pt}{A.~Casimiro}\IEEEauthorrefmark{5},
  \href{mailto:anbessani@fc.ul.pt}{A.~Bessani}\IEEEauthorrefmark{5},
  \href{mailto:jmcecilio@fc.ul.pt}{J.~Cecilio}\IEEEauthorrefmark{5}, 
  \href{mailto:stefan.andersson@veoneer.com}{S.~Andersson}\IEEEauthorrefmark{6},
  \href{mailto:oliver.brunnegard@veoneer.com}{O.~Brunnegard}\IEEEauthorrefmark{6}, 
  \href{mailto:olof.eriksson@veoneer.com}{O.~Eriksson}\IEEEauthorrefmark{6},\\
  \href{mailto:rolandweiss@siemens.com}{R.~Weiss}\IEEEauthorrefmark{7},
  \href{mailto:franz.meierhoefer@siemens.com}{F.~Meierh\"ofer}\IEEEauthorrefmark{7},
  \href{mailto:hans@embedl.ai}{H.~Salomonsson}\IEEEauthorrefmark{8}, \href{mailto:elaheh@embedl.ai}{E.~Malekzadeh}\IEEEauthorrefmark{8},
  \href{mailto:daniel@embedl.ai}{D.~\"Odman}\IEEEauthorrefmark{8},
  \href{mailto:anum.khurshid@ri.se}{A.~Khurshid}\IEEEauthorrefmark{9},\\
  \href{mailto:pascal.felber@unine.ch}{P.~Felber}\IEEEauthorrefmark{10},
  \href{mailto:marcelo.pasin@unine.ch}{M.~Pasin}\IEEEauthorrefmark{10}, 
  \href{mailto:valerio.schiavoni@unine.ch}{V.~Schiavoni}\IEEEauthorrefmark{10},
  \href{mailto:james.menetrey@unine.ch}{J.~M\'en\'etrey}\IEEEauthorrefmark{10},
  \href{mailto:kgugala@antmicro.com}{K.~Gugala}\IEEEauthorrefmark{11},
  \href{mailto:pzierhoffer@antmicro.com}{P.~Zierhoffer}\IEEEauthorrefmark{11},
  \href{mailto:eric.knauss@cse.gu.se}{E.~Knauss}\IEEEauthorrefmark{12},
  \href{mailto:hans-martin.heyn@gu.se}{H.~Heyn}\IEEEauthorrefmark{12}
  }
  \IEEEauthorblockA{\IEEEauthorrefmark{1}\href{https://www.cit-ec.de/en/ks}{Bielefeld University, Germany} |
    \IEEEauthorrefmark{2}\href{https://christmann.info}{christmann informationstechnik + medien GmbH \& Co. KG, Germany} \\
    \IEEEauthorrefmark{3}\href{https://www.inf.uni-osnabrueck.de/research_groups/computer_engineering.html}{Osnabr\"uck University, Germany} | 
    \IEEEauthorrefmark{4}\href{https://www.chalmers.se}{Chalmers University of Technology, Sweden} | 
    \IEEEauthorrefmark{5}\href{http://www.di.fc.ul.pt/~casim}{University of Lisbon, Portugal} \\
    \IEEEauthorrefmark{6}\href{https://www.veoneer.com}{VEONEER Inc., Sweden} | 
    \IEEEauthorrefmark{7}\href{https://www.siemens.com}{Siemens AG, Germany} |
    \IEEEauthorrefmark{8}\href{https://embedl.ai}{EMBEDL AB, Sweden}  |
    \IEEEauthorrefmark{12}\href{https://www.gu.se/en}{G\"oteborg University, Sweden}  \\
    \IEEEauthorrefmark{9}\href{https://www.ri.se/en}{Research Institutes of Sweden AB (RISE)} | 
    \IEEEauthorrefmark{10}\href{http://www.unine.ch/iiun/home/chaires-de-recherche/systemes-complexes.html}{University of Neuch\^{a}tel, Switzerland} | 
    \IEEEauthorrefmark{11}\href{https://antmicro.com}{Antmicro, Poland} 
  }
}

\maketitle

\def\confname{25th Conference \& Exhibition on Design, Automation \& Test in Europe (DATE'22)}
\def\confyear{2022}
\def\confdoi{10.23919/DATE54114.2022.9774653}

\definecolor{yellowPaper}{HTML}{fff8ae}
\AddToShipoutPictureFG*{\AtTextUpperLeft{\begin{tcolorbox}[width=\textwidth,colback=yellowPaper,enhanced,frame hidden,sharp corners]  
        \centering\scriptsize
        \copyright~\confyear\ IEEE. Personal use of this material is permitted. Permission from IEEE must be obtained for all other uses, in any current or future media, including reprinting/republishing this material for advertising or promotional purposes, creating new collective works, for resale or redistribution to servers or lists, or reuse of any copyrighted component of this work in other works.
        This is the author's version of the work. The definitive version has been published in the proceedings of the\\
        \confname.
        \href{https://doi.org/\confdoi}{DOI: \confdoi}
     \end{tcolorbox}   
  }}

\hypersetup{
    pdfcopyright={\copyright~\confyear\ IEEE. Personal use of this material is permitted. Permission from IEEE must be obtained for all other uses, in any current or future media, including reprinting/republishing this material for advertising or promotional purposes, creating new collective works, for resale or redistribution to servers or lists, or reuse of any copyrighted component of this work in other works.
    This is the author's version of the work. The definitive version has been published in the proceedings of the \confname.}
}
 
\begin{abstract}
The VEDLIoT project targets the development of energy-efficient Deep Learning for distributed AIoT applications. A holistic approach is used to optimize algorithms while also dealing with safety and security challenges. The approach is based on a modular and scalable cognitive IoT hardware platform. Using modular microserver technology enables the user to configure the hardware to satisfy a wide range of applications. VEDLIoT offers a complete design flow for Next-Generation IoT devices required for collaboratively solving complex Deep Learning applications across distributed systems. The methods are tested on various use-cases ranging from Smart Home to Automotive and Industrial IoT appliances. VEDLIoT is an H2020 EU project which started in November 2020. It is currently in an intermediate stage with the first results available.
\end{abstract} 
\section{The VEDLIoT Approach}

\label{sec:introduction}
\begin{figure*}[!t]
\centering
\includegraphics[width=0.9\textwidth]{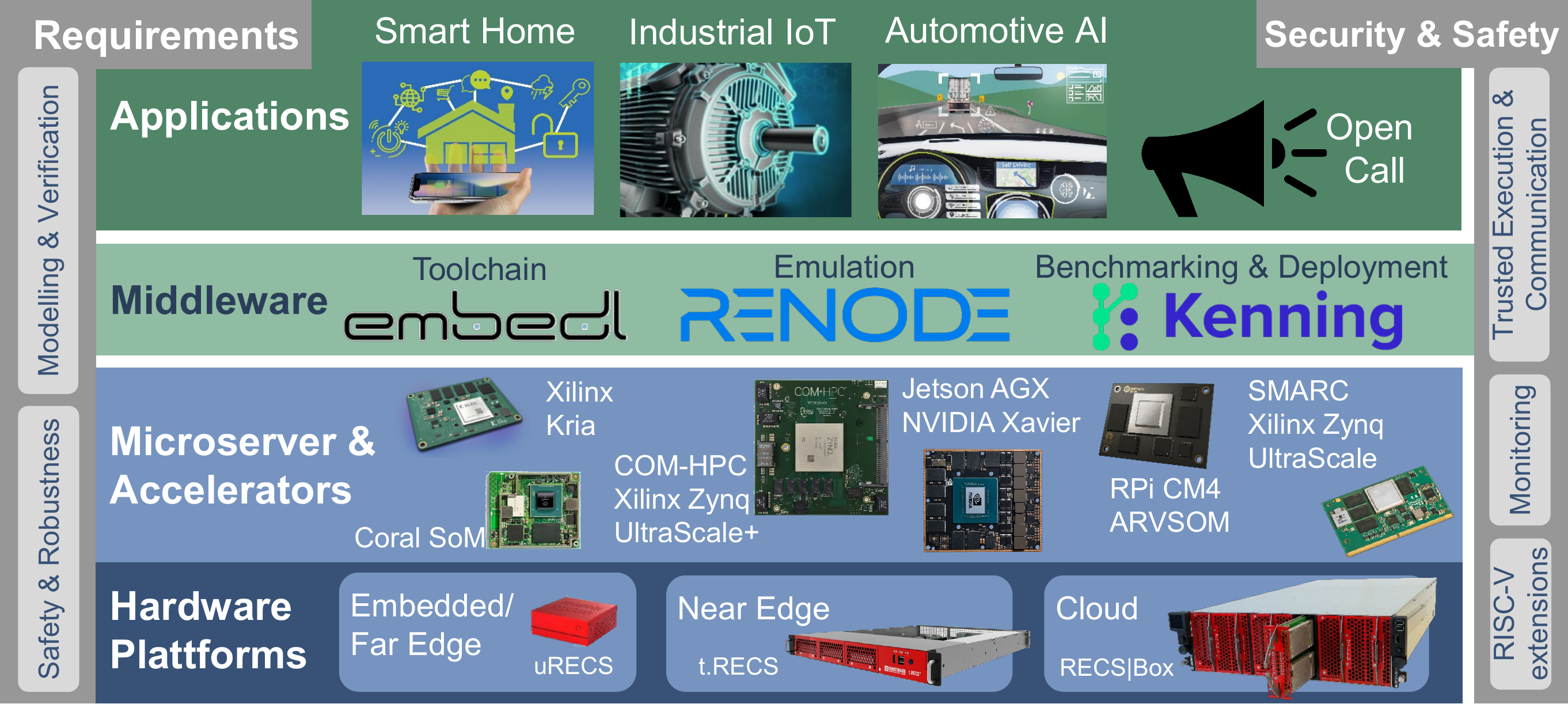} 
\caption{VEDLIoT architecture overview}
\label{fig:vedliot_overview}
\vspace{-1.5em}
\end{figure*}

Deep Learning has become a strong driver in IoT applications. Typically, those applications have very challenging computational demands coupled with a low energy budget. The goal of VEDLIoT is to integrate IoT with Deep Learning, accelerate applications, and optimize them towards energy efficiency. \autoref{fig:vedliot_overview} shows the architecture of VEDLIoT. The project is presented following a bottom-up approach, starting with customizable hardware platforms using modular microservers and specialized hardware accelerators for heterogeneous computing. The optimization of use cases is systematically accompanied from an early stage of development by using requirement engineering and verification techniques for AIoT, also developed within VEDLIoT. Expert-level knowledge of different domains is combined to create a powerful middleware for optimizing the underlying neural networks of deep learning algorithms and ease the development with frameworks for testing, benchmarking and deployment. A lot of development work goes into guaranteeing a high level of safety and security, which is essential for the VEDLIoT use cases.
 \section{Accelerated AIoT hardware platform}
\label{sec:hwplatforms}
The hardware development within VEDLIoT focuses on extending and refining the already available platforms RECS{\textbar}Box and t.RECS, which primary targets cloud and near edge computing. uRECS is developed from scratch within VEDLIoT and focuses on compact dimensions, low cost, and high energy efficiency to better suit low-cost / low-power devices for AI and ML applications (\autoref{fig:vedliot_overview}). Using the RECS hardware platform, VEDLIoT covers the complete range from embedded via edge to cloud computing.

\subsection{Heterogeneous hardware platform}
All RECS hardware platforms share a modular approach, which leads to a heterogeneous, adaptable hardware architecture supporting a wide range of applications and allowing for a future-proof design by an exchangeable/upgradable hardware basis~\cite{griessl2014scalable} and~\cite{legato-date}. Another common feature is the scalable communication-driven infrastructure, realizing efficient communication between heterogeneous microservers via 1\,G/ 10\,G Ethernet and high-speed low-latency connections, reconfigurable during run-time~\cite{oleksiak2017m2dc}.

\begin{figure}[!b]
\centering
\includegraphics[width=0.48\textwidth]{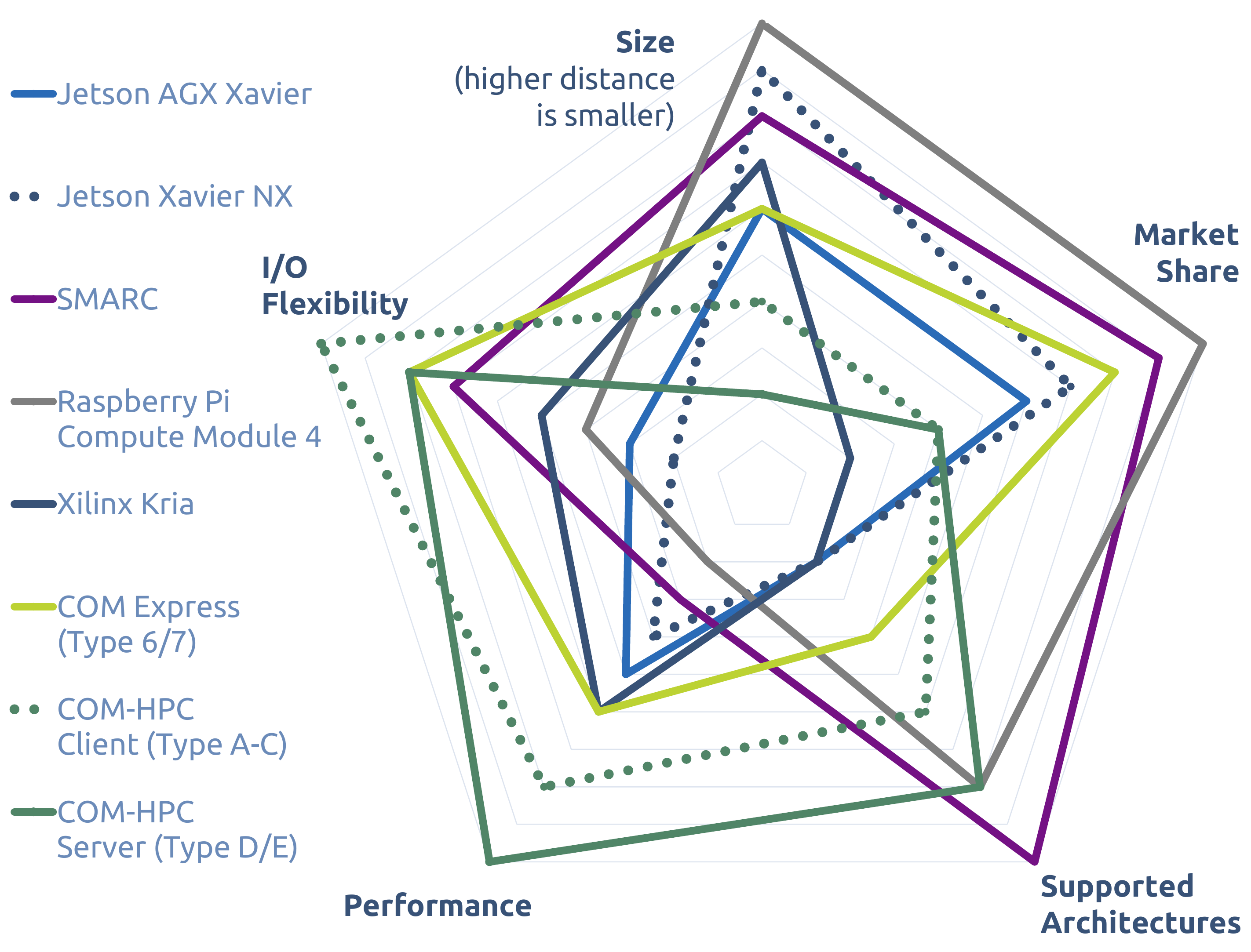} 
\caption{Computer-On-Module (COM) form factors supported by VEDLIoT hardware platforms}
\label{fig:COM_spider}
\vspace{-1.5em}
\end{figure}

Most supported microservers are based on mid- or high-performance Computer-on-Module (COM) form factors, e.g., RECS{\textbar}Box supports COM Express microservers and t.RECS the recently released COM-HPC Server and Client standards.
As shown in \autoref{fig:COM_spider}, several other, well-established form factors focus on low-power embedded computing. SMARC modules, for example, provide a smaller footprint and support with x86, ARM and FPGA-SoC more target architectures. 

uRECS closes the gap in hardware platforms towards embedded/ far edge computing with a power consumption of less than \SI{15}{\watt} as required by some use cases. 
Next to SMARC microservers, it also natively supports Jetson Xavier NX modules. By using adaptor-PCBs, uRECS also integrates Xilinx Kria, and Raspberry Pi compute modules. Extension slots based on USB and M.2 can be used to use additional hardware accelerators or peripherals.

VEDLIoT extends the classically static hardware architecture towards a dynamically configurable infrastructure for increased resource-efficiency and robustness.  The RECS ecosystem enables easy exchange of computing resources and seamless switching between the different heterogeneous components on the system level. On the communication level, e.g., the networking topology or protocol parameters can be adapted to cope with changing real-time or bandwidth requirements. Finally, reconfigurable devices (FPGAs) are utilized to enable the integration of new architectural concepts developed in VEDLIoT. On this level, partial reconfiguration is used to adapt to changing application requirements at run-time, e.g., using implementations with different power/performance footprints.

\subsection{Accelerators and Microservers}
\label{sec:hwplatforms:acc}

\begin{figure*}[!t]
\centering
\includegraphics[width=0.99\textwidth]{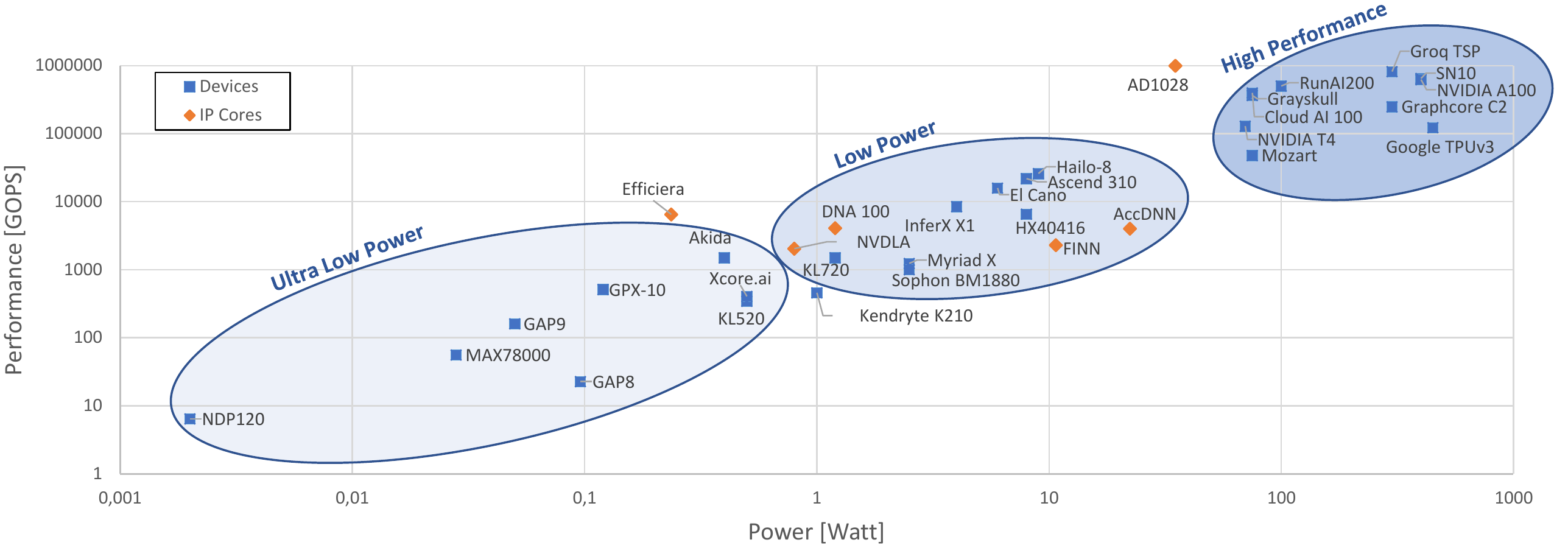} 
\caption{Peak Performance of DL Accelerators}
\label{fig:hardware_mla_perf}
\vspace{-1.5em}
\end{figure*}
One of the key components to delivering the required performance for the Deep Learning (DL) applications are the hardware accelerators. VEDLIoT focuses on developing new dedicated hardware accelerators tailored explicitly towards specific applications requirements. However, the software abstraction layers that have helped the independent development of both software and hardware in the past cannot be used any longer to achieve the best performance and efficiency for the most demanding workloads. The solution is to focus on hardware-software co-design~\cite{codesign}. In VEDLIoT, four different types of DL accelerators are explored: (1) existing off-the-shelf; (2) statically configured; (3) dynamically reconfigurable; and (4) fully simultaneous co-design accelerator. Evaluation and exploration are done using existing accelerators for fast deployment of the required performance. FPGAs are used to develop accelerator prototypes that can achieve higher performance and efficiency for use-case applications. In addition, different modes of operation are identified that offer dedicated accelerators to those modes of operation. These accelerators follow the partial co-design principles by mapping the DL models into the hardware components.
Nevertheless, preliminary results have shown that no single accelerator can provide a better match to different models. Consequently, the fully simultaneous co-design is explored where the hardware is developed together with the software. In addition to mapping the models to hardware, feedback is given to the models so that optimizations can be tuned for better hardware utilization. In addition to the accelerator design, an in-depth study of how the memory is utilized in current accelerators and exploring new approaches for the memory hierarchy for future DL accelerators is performed.
\par
With the many moving parts in the space of DL processing acceleration, VEDLIoT uses Renode, an open-source simulation framework \cite{renode}, to test the FPGA accelerator prototypes. Renode, a functional simulator for complex heterogeneous systems, provides an ability to simulate complete SoCs and run the same software that would be used on hardware. VEDLIoT benefits from Renode’s testing and introspection capabilities, using it both for interactive development of accelerator prototypes and within a Continuous Integration environment. This does not only ease the development process but also makes the final result more reliable.
During the course of the project, Renode is enhanced with capabilities of simulating Custom Function Units, or CFUs. A CFU is an accelerator tightly coupled with the CPU, providing functionality explicitly designed for the planned ML workflow. Programmed in a Hardware Description Language, CFUs are used as an input for Renode to extend simulated cores.

\subsection{Performance Evaluation}
A wide variety of hardware accelerators for deep learning is emerging on the market, targeting a wide range of applications from small embedded systems with power budgets in the order of milliwatt to cloud platforms with a power consumption exceeding 400~W. Figure~\ref{fig:hardware_mla_perf} summarizes the accelerators, that have been analyzed in detail in~\cite{vedliot_d31}.
It has to be noted that the diagram shows a very high-level view. The data is based on the peak performance values (in Giga Operations per Second), provided by the vendors. No normalization to a specific technology node is performed, and the architectures vary in the used precision, ranging from FP32 to INT8 and even binary weights are included. Nevertheless, an interesting fact is that most architectures cluster around an energy efficiency of about 1 Tera Operation per Ws (1~TOPS/W), independent of their individual performance (or power demand).

\begin{figure}[!b]
\centering
\includegraphics[width=0.48\textwidth]{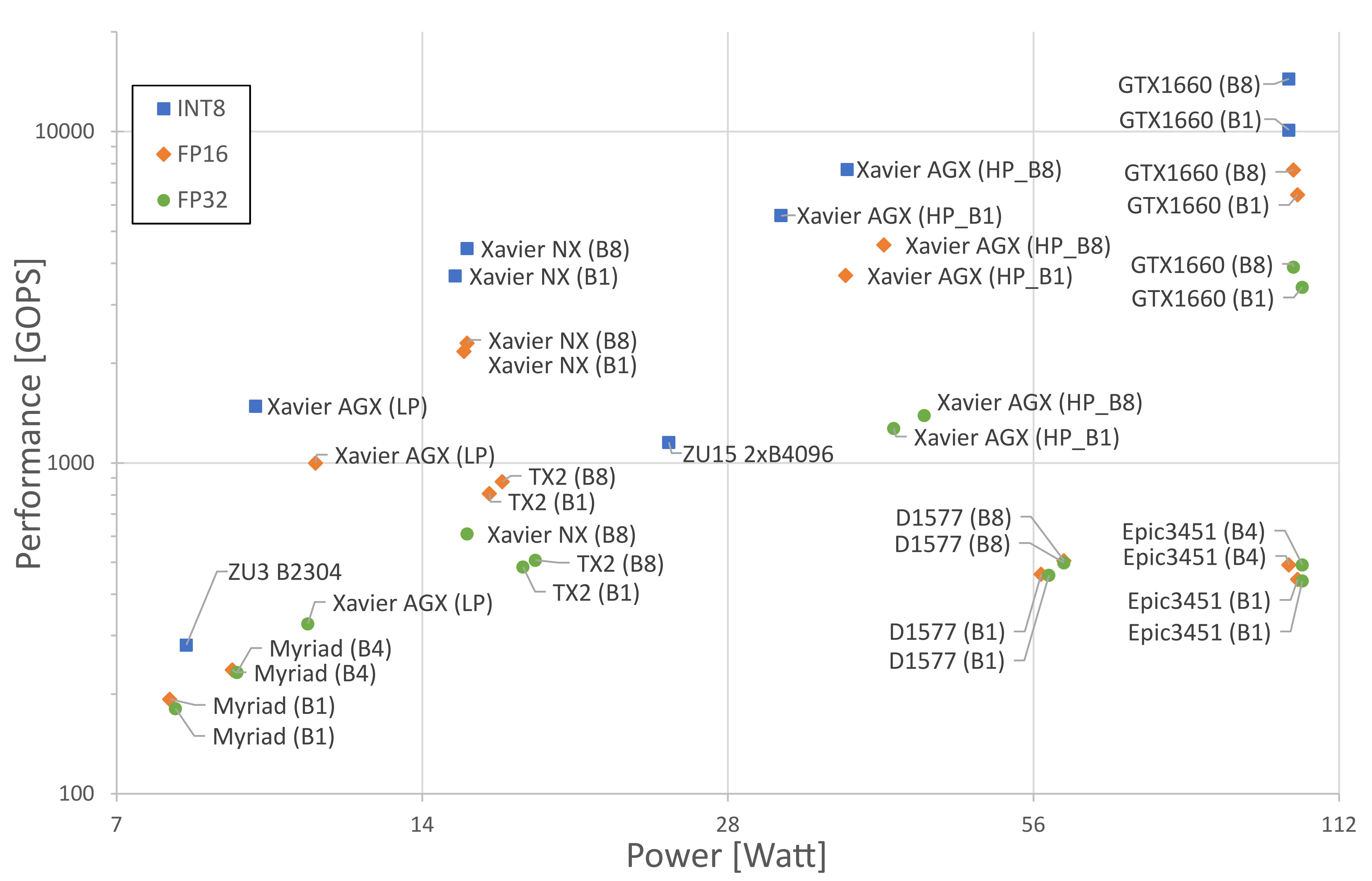} 
\caption{YoloV4 performance evaluation of DL accelerators}
\label{fig:hardware_yolo}
\vspace{-0.5em}
\end{figure}

For performance evaluation, the DL models ResNet50, MobileNetV3 and YoloV4 were chosen to determine comparable performance values of available DL accelerators. Depending on the supported quantization of the hardware, the tests were executed using INT8, FP16 or FP32 datatypes. The tools used for best utilization are chosen based on the manufacturer's recommendations, e.g., TensorRT for NVIDIA. In addition, performance and hardware utilization were optimized by varying the batch size from 1 to 8, which is represented in Figure~\ref{fig:hardware_yolo} by B1, B4 and B8. In this figure, the performance (in GOPS) and the measured power consumption (in Watt) are shown exemplarily for YoloV4. The investigated platforms include x86 CPUs (Epic3451 and D1577), GPUs (GTX1660), eGPUs (Xavier AGX (in high performance and low power mode), Xavier NX and Jetson TX2), FPGAs (Zynq ZU15 and ZU3) and ASICs (Myriad). In VEDLIoT, performance and energy efficiency evaluations are an important basis for selecting DL accelerators to be integrated into the RECS platform, tailoring it towards the use cases.
 \section{Optimizing tool chain for heterogeneous hardware}
\label{sec:toolchain}

Trained deep learning models have redundancy in their computational graph that can be exploited for optimizations. In some cases, models have been compressed down to 49x of their original size, with negligible accuracy loss. This can be achieved by combining methods that remove connections and/or neurons, quantize parameters and activations and encode the parameters in a more compact form \cite{han2015deep}. Although there has been much recent research in the area, most of the results are theoretical speed-ups based on metrics, e.g. number of operations and reduction of parameters. The theoretical speed-ups do not always translate to more efficient execution in hardware \cite{tan2021efficientnetv2}. In the VEDLIoT project, novel methods for hardware-aware optimization are developed.
Furthermore, the industry-standard ONNX, which is an open format to represent machine learning models \cite{onnx}, is used as input to ensure compatibility with the current open ecosystem. The model's computational graph undergoes significant surgery in the optimization phase to optimize its execution latency, power consumption and/or memory footprint. Utilizing the knowledge of the target hardware leads to optimizations that translate to improved execution metrics when deployed. 
Deploying deep learning models on edge devices usually involves the following steps: (1) Preparation and analysis of the dataset, preparation of data pre-processing and output post-processing routines. (2) Model training (usually transfer learning), if necessary. (3) Evaluation and improvement of the model until its quality is satisfactory. (4) Model optimization, usually hardware-specific optimizations (e.g., operator fusion, quantization, neuron-wise or connection-wise pruning). (5) Model compilation to a given target. (6) Model deployment and execution on a given target. There are many different frameworks for most of the above steps (training, optimization, compilation and runtime). The cooperation between those frameworks differs and may provide different results.
\par
Kenning \cite{kenning}, an open-source framework developed by Antmicro, addresses the problem of enabling the tools to cooperate with each other. The interoperability is achieved by converting the models into a common representation using the Open Neural Network Exchange (ONNX) format. All intermediate conversions and optimizations are performed on ONNX models. At the final stage, Kenning converts the model to a selected neural network runtime and deploys it on the target hardware. Based on the implemented interfaces, the Kenning framework can measure the inference duration, resource usage, and processing quality on a given target. Depending on a target, Kenning can monitor inference time,  mean CPU usage, and CPU and GPU memory usage. Kenning can automatically benchmark the processing quality of a given neural network mode and generate a confusion matrix for classification models and recall/precision graphs for detection algorithms. In addition to Kenning, VEDLIoT uses Renode, an open-source simulation framework \cite{renode}, which has been introduced in Section~\ref{sec:hwplatforms:acc}.
 \section{Safety, Security and Requirements for distributed AIoT systems}
\label{sec:safe_sec}
When combining deep learning with the properties of IoT, new concerns might arise that are not yet foreseen by standards and literature. The new concerns include aspects such as data quality, heuristic deep learning modelling, learning of the models, or even new ethical considerations. Additional stakeholders such as data engineers enter the stage, and common languages or interfaces need to be found between the different stakeholders. Typical architectural frameworks, such as the ISO 42010 \cite{ISO42010} or the IEEE 2413 \cite{IEEE2413} standard, cannot cope with concerns for systems that include some form of machine learning. One major challenge identified in VEDLIoT is the difficulty to keep track of dependencies, e.g. through correspondence rules, between the different architectural views. Another problem of current architectural frameworks is the lack of a clear system development hierarchy, which would support the early identification and mapping of dependencies between different architectural views \cite{Nuseibeh2001}.
\subsection{Requirements concepts for AIoT}

Designing a large, distributed system is a hierarchical process \cite{Murugesan2019}. The architectural framework for VEDLIoT not only supports the seamless design and integration of traditional software components and deep learning components but also allow for all necessary quality concerns to be taken into account as early as possible in the design process. The VEDLIoT architectural framework is organized by two aspects: Clusters of concerns, and level of abstraction. These aspects form a 2-dimensional grid of architectural views that guide the concept and design of a VEDLIoT system. Typical clusters of concerns for a system with deep learning components are \textit{logical behavior, process behavior, context and constraints, learning setting, deep learning model, hardware, information, communication, ethical concerns, safety, security, privacy,} and \textit{energy}. Levels of abstraction are \textit{knowledge level, conceptual level, design level}, and \textit{run-time level}. Each architectural view is categorized by the two aspects cluster of concern and level of abstraction. In VEDLIoT, it is shown that dependencies between the architectural views only exist vertically between the views of the same cluster of concern or horizontally between architectural views on the same level of abstraction. This reduces the complexity of the system design challenge and allows for better traceability. Knowledge can become available on all levels of the architectural framework at any time. Traditionally, requirement engineering would be organized in a top-down fashion. However, the architectural framework supports middle-out systems engineering, which is a widely common practice, combining traditional top-down systems design with integration of designated lower-level hardware, software, AI models, or other components \cite{Davis2019}.

\subsection{Safety aspects}
\label{sec:safe_sec:safe}
Safety requirements must guide any model construction for systems using deep neural networks. Safety standards emphasize processes for software development that help with avoiding systematic mistakes during the design of systems. However, safety standards that base on the EN-IEC~61508 standard\footnote{Functional Safety of Electrical/Electronic/Programmable Electronic Safety-related Systems}, such as ISO~26262 for the automotive industry, assume that for software, only systematic faults exist. However, due to the probabilistic nature of deep learning, the assumption that only systematic faults exist in software does not hold anymore. There is no absolute guarantee that a deep neural network performs as intended under the desired circumstances. The desired behavior of the deep neural network depends on the data used for training and validation. Therefore, safety standards for deep neural networks must encompass the deep neural networks and especially the data used for training and validation.
\par
From the perspective of defining an architecture and implementing system solutions to increase the robustness of deep learning processes, VEDLIoT focuses on monitoring approaches to detect faulty situations and trigger appropriate reactive measures. The work is being developed in two directions. Firstly,  the problem of characterizing the quality of the input data is considered, detecting situations in which these data may have been accidentally or even maliciously compromised. A large set of data errors may be easily identified, may be corrected, or the affected data may be removed to avoid the propagation of these errors through the DL models. Different monitoring and error detection mechanisms are developed, depending on the kinds of input data (e.g., time series, image) and on the error types (e.g., outliers, image noise). Secondly, the problem of detecting errors on the output data is analyzed when these errors derive from systematic faults affecting the execution of DL models on devices or edge nodes. It is considered that these faults may have been triggered or injected during run-time (e.g., hardware faults, attacks). In brief, the approach consists in periodically submitting both the input and the output data to a robustness service, which holds a copy of the DL model and can verify the correctness of the output data. To support all these monitors and monitoring mechanisms, an architectural pattern comprising two separate parts is considered, based on the concept of architectural hybridization~\cite{Casimiro:14a}. 
\subsection{Security considerations}
VEDLIoT implements several hardware- and system-level tools to improve the dependability and security of edge applications. Hardware features are leveraged for trusted execution environments combined with well-established dependability techniques to support both middleware and applications layers of the project. So far, the project has focused on developing end-to-end trust through a distributed attestation mechanism, secure execution and communication of critical code (e.g. for monitors, see Section~\ref{sec:safe_sec:safe}) on edge devices. The hardware protection offered by Intel SGX enclaves is leveraged, and an open-source WebAssembly runtime implementation to build a trusted runtime environment without dealing with language-specific APIs. An evaluation shows that SQLite can be fully executed inside an SGX enclave via WebAssembly and existing system interface, with small performance overheads~\cite{DBLP:conf/icde/MenetreyPFS21}.
\par
As part of developing a hardware-trusted execution environment, a novel Trusted Execution Environment (TEE) support for VexRISC-V, an open-source RISC-V soft processor, has been developed. The implementation takes the form of a highly optimized RISC-V Physical Memory Protection (PMP) unit that enables secure processing by limiting the physical addresses accessible by software running on a processor. The PMP unit is configurable in the highest privilege level (the machine mode) and can be used to specify read, write and execute access privileges for a specific memory region. In small devices that only support machine mode (M-mode) and user mode (U-mode), the PMP configurations can efficiently ensure the secure execution of software in M-mode and U-mode. The PMP implementation is part of the official VexRISC-V implementation, and the source code and documentation are openly available for the research community.
\par
Apart from x86 and RISC-V, also ARM SoCs are considered, using TrustZone as a TEE, combined with the open-source and trusted operating system OP-TEE. TrustZone splits the operating system into two parts: the normal and secure worlds. Trusted applications can only run in the secure world, and the operation necessary to change context between worlds is rather complex and cannot be done at user-level. To implement remote attestation for WebAssembly code running in ARM processors, a TEE specification defining how the trusted environment behaves and how the normal world can interact with the secure world is realized. The implementation is based on a root-of-trust provided by the hardware and a secure boot mechanism, preventing an attacker from substituting the trusted software.
 \section{VEDLIoT applications}
\label{sec:apps}
VEDLIoT applications focus on both very high energy efficiency and high-security and safety requirements. These requirements are in line with the general need for \enquote{trusted IoT and edge computing platforms} and \enquote{development and deployment of next-generation computing components} as identified in the EU research agenda \cite{eu-iot-strategy}.
\subsection{Automotive}
Amongst the numerous potential problems in the automotive sector, the Pedestrian Automatic Emergency Breaking (PAEB) was chosen as a well-specified example, which can be benchmarked and compared against state-of-the-art systems. The major development goals are the distribution of the deep learning models and the decision making between different on-car systems and edge devices at varying speeds and reliability of mobile networks. Dynamic distributing of sensor data to edge stations is a quite new research topic. It requires quick monitoring of available mobile networks, their speed and latency, available computing resources of the edge devices and a management system that can quickly react to the current situation. The overall goal is to optimize the energy efficiency in total and minimize the on-car energy consumption. Sending raw sensor data via a mobile network to an edge station always implies a high-security risk. Therefore, an integration of VEDLIoT's remote attestation approach is of importance. 
\subsection{Industrial IoT}
VEDLIoT supports two Industrial IoT use cases: Motor Condition Classification and Arc Detection in DC power distribution cabinets. The Motor Condition Classification use case aims to design and build a prototype of a battery-powered ultra-low energy deep learning-driven small box that can be attached to large electric asynchronous motors and continuously monitors the motor. The states to monitor are the operational, thermal and mechanical conditions of the motor, and upon specified events, e.g. a ball bearing failure, a message is sent to an operator. 
\par
The Arc Detection use case aims to design and build a prototype system that can detect unwanted arcs in DC power distribution cabinets using deep learning technology. A challenge is to guarantee a very low latency from the first spark till inference, including sensing and pre-processing, and an ultra-low false-negative error rate for a smooth operation. In general, arc localization helps for faster fault detection and repair of broken units. 
\subsection{Smart Home}
\begin{figure}[!h]
\centering
\includegraphics[width=0.48\textwidth]{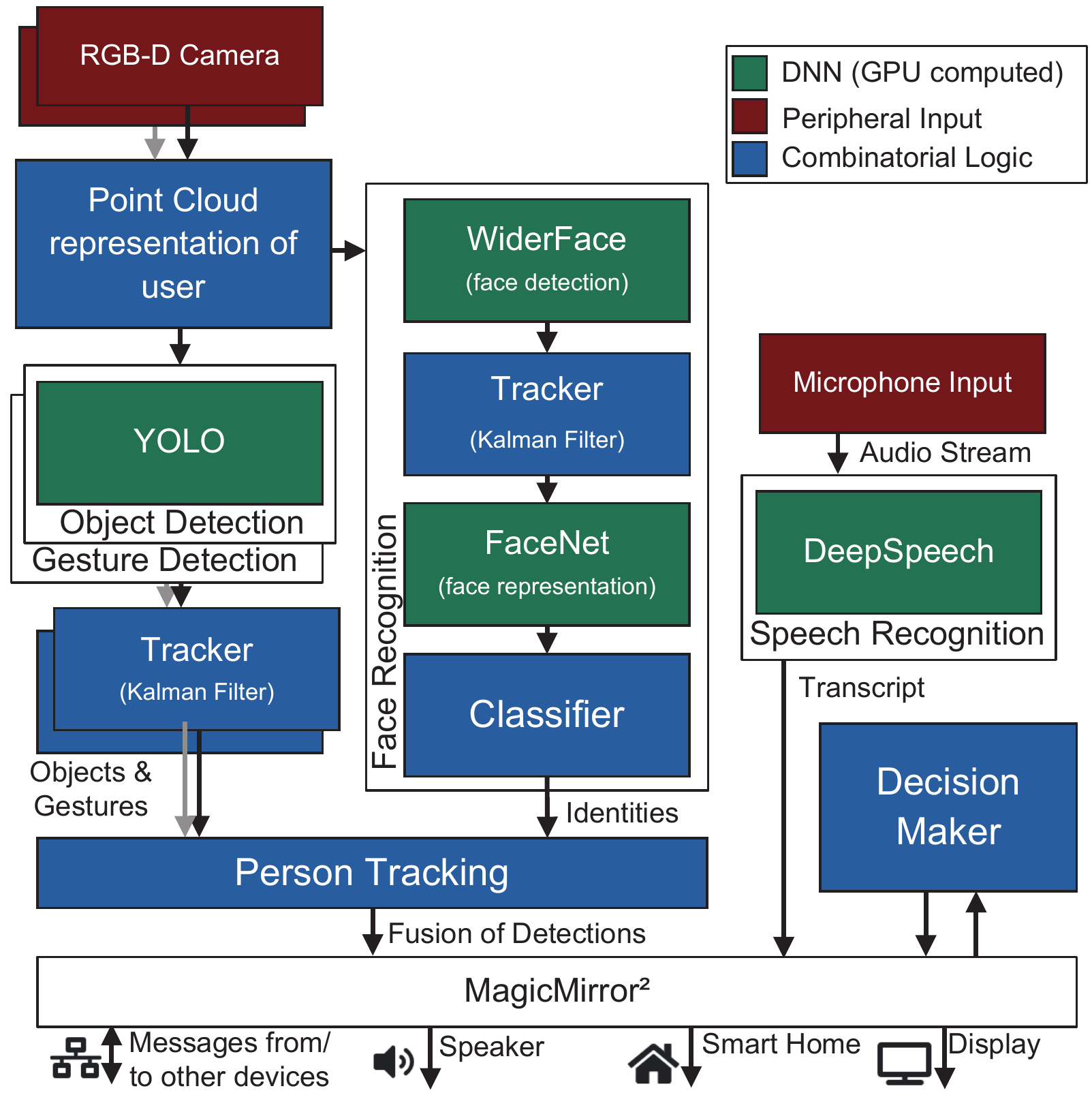}
\caption{Architecture of the Smart Mirror Demonstrator}
\label{fig:SmartMirrorArch}
\vspace{-1.5em}
\end{figure}

This use case targets the development and acceleration of AI-based methods for a demand-oriented interaction between the user and a smart home. To achieve this goal, an intuitive and natural operating interface is crucial, which is realized by a smart-mirror device. As seen in Figure~\ref{fig:SmartMirrorArch}, a camera and a microphone are providing input data, and four different neural networks are used to detect gestures, faces, objects and speech to interact with people. The distribution of data to the cloud is not desirable because of privacy concerns of the residents. Therefore, all sensing and interaction is performed on-site in real-time, making low power and energy efficiency computations a prime concern for this use-case.
 \section{Summary}
\label{sec:summary}
VEDLIoT addresses the challenge of bringing Deep Learning to IoT devices with limited computing performance and low-power budgets. The VEDLIoT AIoT hardware platform provides optimized hardware components and additional accelerators for IoT applications covering the full spectrum from embedded via edge to the cloud. A powerful middleware to ease the programming, test and deployment of neural networks to this heterogeneous hardware. New methodologies for requirement engineering, coupled with safety and security concepts, incorporate the new challenges arising from the use of Deep Learning techniques are designed and applied throughout the complete framework. The concepts are tested driven by challenging use cases in key industry sectors like automotive, automation, and smart home.
\par
In addition, an open call for cascaded funding is foreseen to explore new opportunities by extending the application of the VEDLIoT platform to a more extensive set of new and relevant use cases. Typical open call projects leverage VEDLIoT technologies for their own AI-related IoT use case, thereby broadening the VEDLIoT use-case basis and making the overall concept more robust. The envisaged run-time of the satellite projects is in the range of 9 – 12 months, with an average funding of up to 120,000 € (including 25 \% indirect costs), at a funding/reimbursement rate of 70 \%. More detailed information, including available VEDLIoT technology and the application procedure, is scheduled to be published in early 2022.

\bibliographystyle{unsrt}

\end{document}